\documentclass[aps,pre,showkeys,showpacs,preprint]{revtex4}
\usepackage{graphicx}
\usepackage{amsmath}
\usepackage{amsfonts}
\usepackage{amssymb}
\usepackage{bm}
\usepackage{enumitem}  
\usepackage{float}
\usepackage{color}

\begin{document}
\title{Derivation of Dirac Exchange Interaction Potential from Quantum Plasma Kinetic Theory}

\author{Fernando Haas\footnote{E-mail: fernando.haas@ufrgs.br. Orcid: 0000-0001-8480-6877}}
\affiliation{Instituto de F\'{\i}sica, Universidade Federal do Rio Grande do Sul, Av. Bento Gon\c{c}alves 9500, 91501-970 Porto Alegre, RS, Brasil}

\begin{abstract}
The Dirac exchange interaction is derived from recent quantum kinetic theory for collisionless plasmas. For this purpose, the kinetic equation is written in the semiclassical and long wavelength approximations. The validity of the model for real systems is worked out, in terms of temperature and density parameters. Within the region of applicability, the correlation potential energy is shown to be always smaller than the exchange contribution. From the moments of the quantum kinetic equations, macroscopic, hydrodynamic equations are found, for an electron-ion plasma. The Dirac exchange term is explicitly derived, in the case of a completely degenerate electron gas. These results show, within quantum kinetic theory for charged particle systems, a new view of the Dirac exchange interaction frequently used in density functional theory parametrization. Finally, a simpler form of the quantum plasma exchange kinetic theory is also found. 
\end{abstract}



\maketitle

\section{Introduction}


Exchange effects have been recently included in quantum plasma kinetic theory \cite{Zamanian, Zamanian2}. For this purpose, the complete antisymmetry of the N-particle density matrix was taken into account, which is inline with the Fermi statistics of electrons, in accordance with the Hartree-Fock approximation. For linear waves in completely degenerate plasmas \cite{Ekman}, the results are in exact agreement with older treatments following another methods \cite{Roos}-\cite{Nozieres}.

There has been some discussion about the accordance between the new quantum plasma kinetic theory and the density functional (DFT) modeling, as well as about the adequate quantum hydrodynamics taking into account exchange effects \cite{Ekman}. Due to the analytical and even computational complexity of phase-space models,   
the reduction of variables in averaged, macroscopic theories is an attractive alternative, for both quantum and classical plasma. Exchange (and correlation) effects have been considered e.g. in the recent review \cite{Manfredi} about quantum plasma hydrodynamics. 

The purpose of the present work is to show the complete agreement between exchange quantum kinetic plasma theory \cite{Zamanian, Zamanian2} and the traditional Dirac exchange potential often considered in DFT. This goes a step further than previous treatments where the exchange (and correlation) potential was inserted {\it ab initio} in the Schr\"odinger equation \cite{hh}.

The article is organized as follows. In Section II, we consider the semiclassical limit of the new exchange kinetic theory. This allows considerable simplification of the central kinetic equation. The validity conditions and the approximations made are described in detail. Section III works out the resulting susceptibilities in electron-ion quantum plasmas, for general electrostatic linear waves. With appropriate averaging methods, in Section IV the general exchange fluid equations are derived. Section V shows that in the particular case of a completely degenerate Fermi gas the equation of state is exactly compatible with the Dirac exchange term. Section VI has the conclusions, also pointing the possible generalizations. 

\section{Exchange kinetic contribution in the semiclassical limit}

Assuming the two-particle electron density matrix in the form of an antisymmetric product of two one-particle density matrices, the Pauli principle becomes assured. Following this approach, a kinetic equation for the one-particle Wigner function $f = f({\bf x}, {\bf p}, t)$ can be deduced \cite{Zamanian, Zamanian2}, neglecting spin polarization, quantum diffraction and correlation effects, in the long wavelength approximation, reading 
\begin{eqnarray}
\frac{\partial}{\partial t}\!\!\!&f&\!\!\!({\bf x}, {\bf p}, t) + \!\! \frac{{\bf p}}{m}\cdot\nabla f({\bf x}, {\bf p}, t) + e\nabla\phi({\bf x},t)\cdot\frac{\partial}{\partial{\bf p}} f({\bf x}, {\bf p}, t) \nonumber \\ \label{e1} &=& \frac{\hbar}{2}\frac{\partial}{\partial p_i}\int\!\! d^{3}\rho\, d^{3}q\, e^{-i{\boldsymbol{\rho}}\cdot{\bf q}}\,\frac{\partial V(\boldsymbol{\rho})}{\partial \rho_i}\,f\left({\bf x}-\frac{\hbar\boldsymbol{\rho}}{2}, {\bf p}+\frac{\bf q}{2}, t\right)f\left({\bf x}-\frac{\hbar\boldsymbol{\rho}}{2}, {\bf p}-\frac{\bf q}{2}, t\right)  \\ &-& \!\!\frac{i\hbar^2}{8}\frac{\partial}{\partial p_i}\frac{\partial}{\partial p_j}\!\int\!\! d^{3}
\rho\, d^{3}q\,e^{-i{\boldsymbol{\rho}}\cdot{\bf q}}\,\frac{\partial V(\boldsymbol{\rho})}{\partial \rho_i}\,
\left[f\left({\bf x}\!-\!\frac{\hbar\boldsymbol{\rho}}{2},{\bf p}\!-\!\frac{\bf q}{2},t\right)\!\left(\!\frac{\overleftarrow{\partial}}{\partial x_j} - \frac{\overrightarrow{\partial}}{\partial x_j}\!\right)\!f\left({\bf x}\!-\!\frac{\hbar\boldsymbol{\rho}}{2},{\bf p}\!+\!\frac{\bf q}{2},t\right)\right] . \nonumber 
\end{eqnarray}
Here $\phi = \phi({\bf x},t)$ is the scalar potential and $V = V({\bf x}) = e^{2}/(4\pi\varepsilon_0 |{\bf x}|)$ is the electron-electron Coulomb potential. The summation convention holds and the remaining symbols have their usual meaning while the arrows in some partial derivatives indicate the sense of operation. By construction Eq. (\ref{e1}) takes into account the bare exchange effects, represented by the terms on the right hand side, and can be termed the exchange kinetic equation, which is our starting point. Details are discussed in \cite{Zamanian, Zamanian2}.

To have the exchange kinetic equation in a semiclassical approximation, we have to make a formal series expansion in powers of $\hbar$.
In the long wavelength approximation, the calculation (shown in the Appendix) gives the semiclassical exchange kinetic equation 
\begin{eqnarray}
\frac{\partial}{\partial t}\!\!\!&f&\!\!\!({\bf x}, {\bf p}, t) + \!\! \frac{{\bf p}}{m}\cdot\nabla f({\bf x}, {\bf p}, t) + e\nabla\phi({\bf x},t)\cdot\frac{\partial}{\partial{\bf p}} f({\bf x}, {\bf p}, t) \nonumber \\ \label{e2} &=& \frac{e^2\hbar^2}{2\varepsilon_0}\frac{\partial}{\partial p_i}\frac{\partial}{\partial x_j} \int \frac{d^{3} q}{q^2} \left(\delta_{ij} - \frac{2 q_i q_j}{q^2}\right) f({\bf x},{\bf p} + {\bf q}, t) f({\bf x}, {\bf p} - {\bf q}, t) \,,
\end{eqnarray}
which is considerably simpler than Eq. (\ref{e1}). The last integral on the right hand side of Eq. (\ref{e1}) is of higher order and does not contribute in the semiclassical approximation.

The formal series expansion in powers of $\hbar$ implicitly implies a power series on a dimensionless quantity $\hbar/(m\,v_0\,l_0) \ll 1$, where $v_0$ is a natural velocity scale and $l_0$ is a natural length scale. The same approximation justifies the neglect of quantum diffraction, also known as quantum recoil, which was already assumed for the derivation \cite{Zamanian, Zamanian2}  of the basic kinetic equation (\ref{e1}). The semiclassical approximation keeps the first order exchange effects. It is a welcome avenue, to analyze the implications of such approximation. 


Since exchange effects are more stringent for dense plasmas, we focus on degenerate plasmas. For a degenerate plasma, a natural velocity scale is the Fermi velocity $v_F$, so that $v_0 = v_F$. On the other hand, a characteristic length $l_0$ could be given  by the electron 
Thomas-Fermi length $\lambda_F = v_F/\omega_p$ for stationary structures \cite{book}, where $\omega_p$ is the plasma frequency, or by $\lambda = 2\pi/k$ for linear wave propagation with wavenumber $k$. For the sake of definiteness, setting $l_0 = \lambda_F$ and taking into account $v_F = \hbar (3\pi^2 n_0)^{1/3}/m, \,\, \omega_p = [n_0 e^2/(m \varepsilon_0)]^{1/2}$, where $n_0$ is the equilibrium electrons number density, $m$ is the electron mass, $e$ is the elementary charge and $\varepsilon_0$ is the vacuum permittivity, the semiclassical approximation traduces into $a_0 \, n_0^{1/3} \gg 4\pi/(3\pi^2)^{4/3} = 0.14$, where $a_0 = 4\pi\varepsilon_0\hbar^2/(m e^2)$ is the Bohr radius. Hence, for stationary structures in degenerate plasma the semiclassical approximation applies for $n_0 \gg 1.7 \times 10^{28}\,{\rm m}^{-3}$, a condition on the number density only. 

\begin{figure}[h]
\begin{center}
\includegraphics[width=10.5 cm]{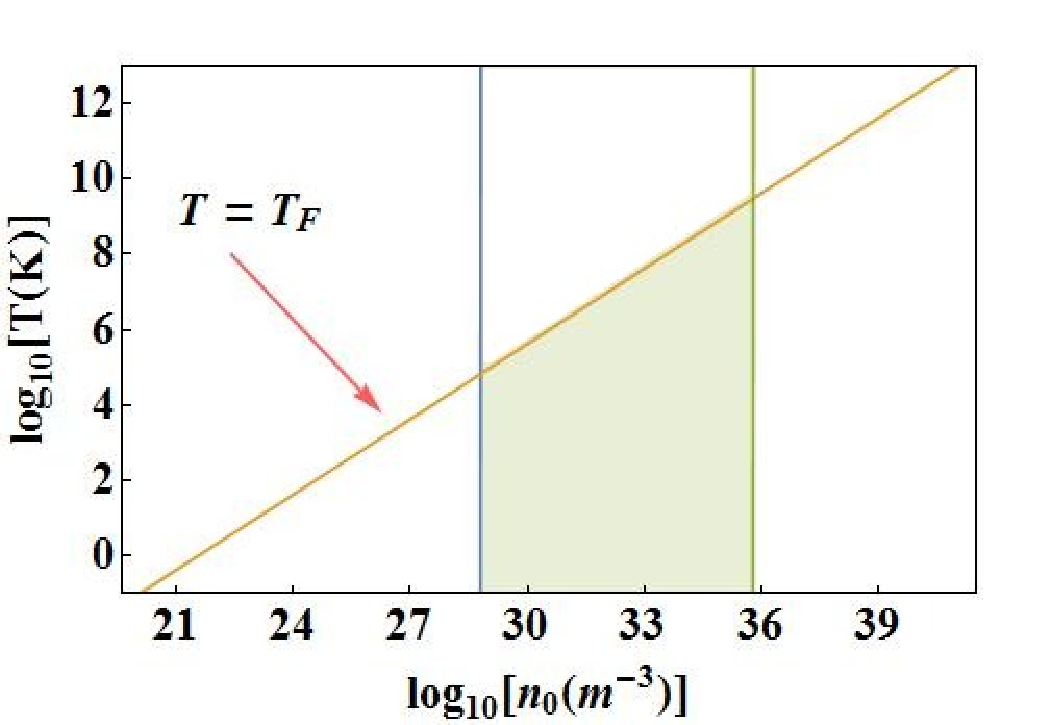}
\caption{Density-temperature diagram for which the present modeling is applicable, in a logarithmic scale, where $n_0$ is the equilibrium number density (measured in ${\rm m}^{-3}$) and $T$ is the thermodynamic temperature in $K$. The left vertical line indicates the minimal number density for which correlation effects are negligible, also justifying the semiclassical approximation. The right vertical line indicates the maximal number density to avoid relativistic effects. Degenerate plasmas are below the straight line $T = T_F$. The modeling is valid in the filled area.}
\label{fig1}
\end{center}
\end{figure}

The remaining validity conditions are as follows. Neglect of correlations imply that the average Coulomb energy $E_C = e^2 n_{0}^{1/3}/(4\pi\varepsilon_0)$ is much smaller than the average kinetic energy, which is the Fermi energy $E_F = m v_F^2/2$ in an order of magnitude estimate. Defining the coupling parameter $g_c = E_C/E_F$ we then find the necessary condition 
\begin{equation}
\label{g}
g_c = \frac{E_C}{E_F} = \frac{2}{(3\pi^2)^{2/3} a_0\,n_{0}^{1/3}} \ll 1 \,,
\end{equation}
which at the end is again a function of the number density only. Equation (\ref{g}) implies $n_0 \gg 6.1 \times 10^{28}\,{\rm m}^{-3}$, which is very similar to the semiclassical condition. Moreover, it is often overlooked that the application of a non-relativistic model is possible only when relativistic effects are negligible. To avoid a relativistic mass increase, one needs at least $v_F/c \ll 1$, where $c$ is the speed of light. This amounts to $n_0 \lambda_C^3 \ll 1/(3\pi^2)$, where $\lambda_C = \hbar/(m c)$ is the Compton length. Therefore, $n_0 \ll 5.9 \times 10^{35}\,{\rm m}^{-3}$, excluding very dense plasmas deserving a relativistic treatment. Finally, the degeneracy condition is $T \ll T_F$, where $T$ is the thermodynamic temperature, $T_F = E_F/\kappa_B$ is the Fermi temperature and $\kappa_B$ is the Boltzmann condition. All in all, the region for which the present modeling is applicable is shown in the filled area in the density-temperature diagram in Fig. \ref{fig1}, in a logarithmic scale.

On the other hand, concerning wave propagation and setting $l_0 = \lambda$, from $\hbar/(m v_F \lambda) \ll 1$ one gets $\lambda n_{0}^{1/3} >> 1/(3\pi^2)^{1/3}$ which means the wavelength should be much larger than the mean inter-particle distance. Therefore the semiclassical approximation is equivalent to a continuous media (fluid) assumption, a long wavelength condition in this case. 

\begin{figure}[h]
\begin{center}
\includegraphics[width=10.5 cm]{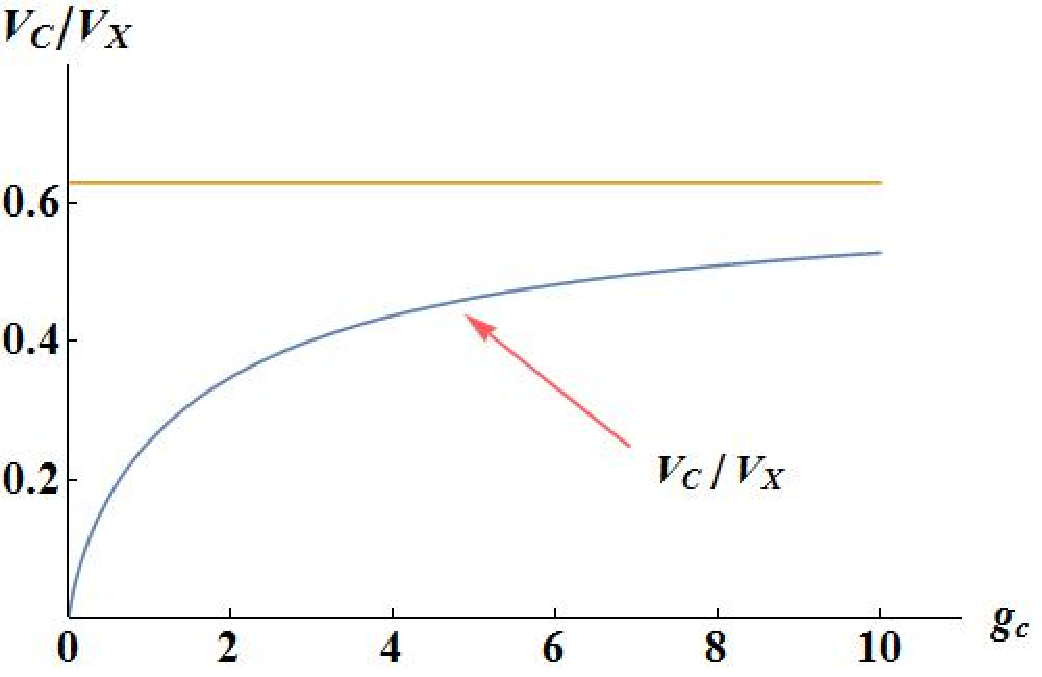}
\caption{Ratio between effective correlation and exchange potentials as a function of the coupling parameter $g_c$, from Eq. (\ref{vcvx}). The horizontal straight line shows the limiting value $V_C/V_X = 0.63$ as $g_C \gg 1$.}
\label{fig2}
\end{center}
\end{figure}

It is interesting to look into the neglect of correlations from the point of view of density-functional theory. The DFT exchange-correlation potential $V_{XC}$ is given \cite{Hedin, Brey} by 
\begin{equation}
V_{XC} = V_X + V_C = g_D \left(\frac{n}{n_0}\right)^{1/3} \left[1 + \frac{0.034}{a_0 n^{1/3}}\,\ln\left(1 + 18.37 a_0 n^{1/3}\right)\right] \,,
\end{equation}
where 
\begin{equation}
V_X = g_D \left(\frac{n}{n_0}\right)^{1/3} \,, \quad g_D = 0.985\, \frac{(3\pi^2)^{2/3}}{4\pi}\frac{\hbar^2 \omega_p^2}{m v_F^2} 
\end{equation}
is the Dirac potential form of the exchange potential \cite{Dirac} and 
\begin{equation}
V_C = V_X \times \left[\frac{0.034}{a_0 n^{1/3}}\,\ln\left(1 + 18.37 a_0 n^{1/3}\right)\right] 
\end{equation}
is the correlation potential. 
Evaluating at the equilibrium number density and using Eq. (\ref{g}), one has 
\begin{equation}
\label{vcvx}
\frac{V_C}{V_X} = 0.16\, g_c\, \ln\left(1 + \frac{3.84}{g_c}\right) \,,
\end{equation}
shown in Fig. \ref{fig2}. As expected, the relative importance of the correlation effects (similar to collisionall effects) goes to zero as the coupling parameter $g_c \ll 1$. On the opposite limit, when $g_c \gg 1$ one has $V_C/V_X = 0.63$, which is a considerable value but still such that $V_C < V_X$, which is thus always true at least from the point of view of the effective DFT exchange-correlation potential. 

\section{Susceptibilities in an Electron-Ion Plasma}

In an electron-ion plasma we have to consider the ions probability distribution function $f_i = f_i({\bf x}, {\bf p}, t)$. It follows a classical Vlasov equation 
\begin{equation}
\frac{\partial}{\partial t} f_i({\bf x}, {\bf p}, t) +  \frac{{\bf p}}{M}\cdot\nabla f_i({\bf x}, {\bf p}, t) - e\nabla\phi({\bf x},t)\cdot\frac{\partial}{\partial{\bf p}} f_i({\bf x}, {\bf p}, t) = 0 \,,
\label{fi}
\end{equation}
due to a presumably larger ion mass $M \gg m$. For simplicity ions have been assumed to be single ionized. To close the system we have Poisson's equation
\begin{equation}
\label{poi} 
\nabla^{2}\phi = \frac{e}{\varepsilon_0}\,\left(\int d^{3}p\, f({\bf x}, {\bf p}, t) - \int d^{3}p\, f_i({\bf x}, {\bf p}, t)\right) \,.
\end{equation}

For small amplitude waves it is assumed plane wave perturbations as
\begin{eqnarray}
f &=& f_{0}({\bf p}) + \delta f({\bf p})\,\exp[i({\bf k}\cdot{\bf x} - \omega t)] \,, \quad  f_i = f_{0(i)}({\bf p}) + \delta f_{i}({\bf p})\,\exp[i({\bf k}\cdot{\bf x} - \omega t)] \,, \nonumber \\
\phi &=& \delta\phi\,\exp[i({\bf k}\cdot{\bf x} - \omega t)] \,,
\end{eqnarray}
where $\delta$ denotes first order quantities.  Linearizing yields the dispersion relation 
\begin{equation}
1 + \chi_{e} + \chi_{i} = 0 \,,
\end{equation}
where the electron-ion susceptibilities $\chi_{e,i}$ are 
\begin{equation}
\chi_e = \frac{e}{\varepsilon_0 k^2}\int d^{3}p \,\frac{\delta f}{\delta\phi} \,, \quad \chi_i = - \,\frac{e}{\varepsilon_0 k^2}\int d^{3}p \, \frac{\delta f_i}{\delta\phi} \,.
\end{equation}

Linearizing the ions kinetic equation (\ref{fi}) and assuming a cold ions equilibrium $f_{i(0)} = n_0\,\delta({\bf p})$ we get $\chi_i = - \omega_{pi}^2/\omega^2$, where $\omega_{pi} = [n_0 e^2/(M\varepsilon_0)]^{1/2}$. For electrons we linearize the exchange kinetic equation (\ref{e2}) to get 
\begin{eqnarray}
\label{fii}
&\strut& \left(-\omega + \frac{{\bf k}\cdot{\bf p}}{m}\right)\,\delta f({\bf p}) + e \,\delta\phi \,{\bf k}\cdot\frac{\partial f_{0}({\bf p})}{\partial{\bf p}}  \\ &\strut& = \frac{e^2 \hbar^2}{2\varepsilon_0}\,\frac{\partial}{\partial p_i} \int \frac{d^{3}q}{q^2}\left(k_i - 2\,{\bf k}\cdot{\bf q}\,\frac{q_i}{q^2}\right) \Bigl(f_{0}({\bf p} + {\bf q}) \,\delta f({\bf p} - {\bf q}) + f_{0}({\bf p} - {\bf q}) \,\delta f({\bf p} + {\bf q})\Bigr) \,. \nonumber 
\end{eqnarray}

At this point we remember that under the semiclassical condition the entire right-hand side of Eq. (\ref{fii}) is itself a correction. Hence it is legitimate to insert the classical expression 
\begin{equation}
\delta f({\bf p}) = \frac{e\,\delta\phi\,{\bf k}\cdot\partial f_{0}({\bf p})/\partial{\bf p}}{\omega - {\bf k}\cdot{\bf p}/m} 
\end{equation}
into the exchange term of Eq. (\ref{fii}). In this way the electrons susceptibility decomposes into classical $\chi_{e}^C$ and exchange $\chi_{e}^X$ contributions such that $\chi_e = \chi_{e}^C + \chi_{e}^{X}$, where 
\begin{eqnarray}
\chi_{e}^C &=& \frac{e^2}{\varepsilon_0 k^2}\int d^{3}p\,\frac{{\bf k}\cdot\partial f_{0}({\bf p})/\partial{\bf p}}{\omega - {\bf k}\cdot{\bf p}/m} \,, \label{xe} \\
&\strut& \nonumber \\
\chi_{e}^X &=& - \,\frac{1}{2}\left(\frac{e^2 \hbar}{\varepsilon_0 k}\right)^2 \int \frac{d^{3}p}{\omega - {\bf k}\cdot{\bf p}/m}\,\frac{\partial}{\partial p_i} \int \frac{d^{3}q}{q^2} \times \nonumber \\
&\times& \left(k_i - 2\,{\bf k}\cdot{\bf q}\,\frac{q_i}{q^2}\right) \left(\frac{f_{0}(+) \,{\bf k}\cdot\partial f_{0}(-)/\partial{\bf p}}{\omega - {\bf k}\cdot({\bf p} - {\bf q})/m} + \frac{f_{0}(-) \,{\bf k}\cdot\partial f_{0}(+)/\partial{\bf p}}{\omega - {\bf k}\cdot({\bf p} + {\bf q})/m}\right) \,, \label{chie}
\end{eqnarray}
in terms of the shorthand $f_{0}(\pm) = f_{0}({\bf p} \pm {\bf q})$. 

It is instructive to keep the full kinetic equation (\ref{e1}) and perform again the previous operations, to find the same $\chi_{e}^C$ as in Eq. (\ref{xe})  and 
\begin{eqnarray}
\chi_{e}^X &=& - \,\frac{2\,\hbar\, e^4}{\varepsilon_{0}^2 k^2} \int \frac{d^{3}p}{\omega - {\bf k}\cdot{\bf p}/m}\,\frac{\partial}{\partial p_i} \int \frac{d^{3}q \,(q_i + \hbar k_i/4)}{|{\bf q} + \hbar{\bf k}/4|^2} \times \nonumber \\
&\times& \left(\frac{f_{0}(+) \,{\bf k}\cdot\partial f_{0}(-)/\partial{\bf p}}{\omega - {\bf k}\cdot({\bf p} - {\bf q})/m} + \frac{f_{0}(-) \,{\bf k}\cdot\partial f_{0}(+)/\partial{\bf p}}{\omega - {\bf k}\cdot({\bf p} + {\bf q})/m}\right) \label{ful}
\end{eqnarray}
instead of Eq. (\ref{chie}). However, expanding taking into account the long wavelength condition $\hbar k \ll m v_0$ as discussed in Section II it is easy to regain Eq. (\ref{chie}) from Eq. (\ref{ful}), as verified from parity properties. In particular it is immediate to verify that these parity properties show that the first non-zero contribution is proportional to $\hbar^2$. 


\section{Fluid Equations}

Macroscopic equations can be derived from the moments of the simplified exchange kinetic equation (\ref{e2}). In this way it is even possible to derive an exchange potential $V_X$ from first principles. In the case of completely degenerate electrons it turns out that $V_X$ is the Dirac exchange interaction potential, as will be shown next. 

As usual \cite{h2021} 
the number density $n = n({\bf x}, t)$, the velocity field ${\bf u} = {\bf u}({\bf x}, t)$ and the pressure dyad ${\bf P} = {\bf P}({\bf x}, t)$ are defined in terms of the moments 
\begin{eqnarray}
n &=& \int d^{3}p\,f \,, \\
m n {\bf u} &=& \int d^{3}p \,{\bf p}\, f \,, \\
{\bf P} &=& \frac{1}{m} \int d^{3}p \,{\bf p} \otimes {\bf p}\, f - m n {\bf u} \otimes {\bf u} \,.
\end{eqnarray}
Higher order moments could be easily implemented as well. 

From Eq. (\ref{e2}) and with appropriate boundary conditions in momentum space we find the continuity equation 
\begin{equation}
\frac{\partial n}{\partial t} + \nabla\cdot(n {\bf u}) = 0 \,, \label{cont} 
\end{equation}
and the momentum balance equation 
\begin{eqnarray}
m\,\left(\frac{\partial}{\partial t} + {\bf u}\cdot\nabla\right) u_i &=& - \,\frac{1}{n}\frac{\partial P_{ij}}{\partial x_j} + e\,\frac{\partial\phi}{\partial x_i} \label{cw} \\ 
&\strut& - \,\frac{e^2 \hbar^2}{2\,\varepsilon_0 n}\frac{\partial}{\partial x_j} \int \frac{d^{3}p\, d^{3}q}{q^2} \left(\delta_{ij} - \frac{2\,q_i q_j}{q^2}\right) \Bigl(f({\bf x}, {\bf p} + {\bf q}) f({\bf x}, {\bf p} - {\bf q})\Bigr) \,, \nonumber 
\end{eqnarray}
expressed in component-wise manner, where the explicit time-dependence was omitted in $f$, for brevity. The momentum balance equation is valid for any local $f({\bf x}, {\bf p})$. 

Equation (\ref{cw}) can be written as 
\begin{equation}
\label{n7}
m\,\left(\frac{\partial}{\partial t} + {\bf u}\cdot\nabla\right) {\bf u} = - \,\frac{\nabla\cdot{\bf P}}{n} - \,\frac{\nabla\cdot{\bf P}^{X}}{n} + e\,\nabla\phi \,,
\end{equation}
in terms of the exchange pressure dyad ${\bf P}^{X}$ defined by 
\begin{equation} 
P_{ij}^{X} = \,\frac{e^2 \hbar^2}{2\,\varepsilon_0} \int \frac{d^{3}p\, d^{3}q}{q^2} \left(\delta_{ij} - \frac{2\,q_i q_j}{q^2}\right) \Bigl(f({\bf x}, {\bf p} + {\bf q}) f({\bf x}, {\bf p} - {\bf q})\Bigr) \,, 
\end{equation}
in component-wise manner. The simplicity of Eq. (\ref{n7}) in comparison with Eq. (7) of Ref. \cite{h2021} comes from starting from Eq. (\ref{e2}) instead of the full exchange kinetic equation (\ref{e1}), which involves no loss of generality as long as the validity conditions discussed in Section II remain valid. 

For isotropic in momentum distributions, namely if $f = f({\bf x}, p), p = |{\bf p}|$, it follows that $P_{ij} = p\,\delta_{ij}$ and $P_{ij}^X = p_X \delta_{ij}$, where $p$ is the scalar pressure $p = (1/3) {\rm Tr} P_{ij}$ (${\rm Tr}$ denotes the trace) and where $p_X$ is the exchange scalar pressure, 
\begin{equation}
\label{xp} 
p_X = \frac{1}{3} {\rm Tr} P_{ij}^X = \frac{e^2 \hbar^2}{6\,\varepsilon_0} \int \frac{d^{3}p\, d^{3}q}{q^2} \Bigl(f({\bf x}, {\bf p} + {\bf q}) f({\bf x}, {\bf p} - {\bf q})\Bigr) \,.
\end{equation}

Similarly, for ions one can define 
\begin{eqnarray}
n_i &=& \int d^{3}p\,f_i \,, \\
M n_i {\bf u}_i &=& \int d^{3}p \,{\bf p}\, f_i \,, \\
{\bf P}_i &=& \frac{1}{M} \int d^{3}p \,{\bf p} \otimes {\bf p}\, f_i - M n_i {\bf u}_i \otimes {\bf u}_i \,.
\end{eqnarray}

The moments of the Vlasov equation (\ref{fi}) yield 
\begin{eqnarray}
\frac{\partial n_i}{\partial t} + \nabla\cdot(n_i {\bf u}_i) &=& 0 \,, \label{conti}  \\
\label{ni7}
M\,\left(\frac{\partial}{\partial t} + {\bf u}_i\cdot\nabla\right) {\bf u}_i &=& - \,\frac{\nabla\cdot{\bf P}_i}{n_i} - e\,\nabla\phi \,. 
\end{eqnarray}
Finally one has Poisson's equation
\begin{equation}
\label{poii} 
\nabla^{2}\phi = \frac{e}{\varepsilon_0}\,(n - n_i) \,.
\end{equation}
To have closure of the system, one needs to express all pressure dyads in terms of lower order moments. This can be achieved assuming a local equilibrium distribution function, as shown in the next Section. 

\section{Closure for Completely Degenerate Electrons}

For a completely degenerate electron gas, one has 
\begin{equation}
\label{dfd}
f({\bf x}, {\bf p}) = A\,\theta(\tilde{p}_F - |{\bf p} - m {\bf u}|) \,, \quad A = \frac{3\,n}{4\pi \tilde{p}_{F}^{3}} \,,
\end{equation}
where $\theta$ is the Heaviside step function of the indicated argument and where $\tilde{p}_F = \hbar\,(3\pi^2 n)^{1/3}$ is the local Fermi momentum depending on the local number density $n$. Equation (\ref{dfd}) represents a local, velocity displaced zero temperature Fermi-Dirac distribution. 

In the reference frame of the electrons fluid, the quasi-equilibrium (\ref{dfd}) is isotropic in momentum space. It is immediate to evaluate the scalar pressure 
\begin{equation}
p = \frac{2}{5}\,n_0 E_F \left(\frac{n}{n_0}\right)^{5/3} \,, \quad E_F = \frac{p_{F}^2}{2\,m} \,, \quad p_F = \hbar\,(3\pi^2 n_0)^{1/3} \,,
\end{equation}
which is the equation of state for the completely degenerate electron gas where $E_F, p_F$ are the Fermi energy and momentum, and $n_0$ is the equilibrium number density. 

The derivation of the equation of state for the exchange scalar pressure (\ref{xp}) involves the evaluation of the volume intersection between two spheres of identical radius $\tilde{p}_F$, centered at ${\bf p} = \pm {\bf q}$, which is 
\begin{equation}
\int d^{3}p\, \theta(\tilde{p}_F - |{\bf p} + {\bf q}|)\,\theta(\tilde{p}_F - |{\bf p} - {\bf q}|) = \frac{4\pi}{3}\,(\tilde{p}_F + \frac{q}{2})\,(\tilde{p}_F - q)^2\,\theta(\tilde{p}_F - q) \,,
\end{equation}
as found from elementary calculus \cite{Kern}. The remaining integration in Eq. (\ref{xp}) gives 
\begin{equation}
\label{px}
p_X = \frac{3}{16}\,\frac{e^2}{\varepsilon_0}\,\frac{n^{4/3}}{(3\pi^2)^{2/3}} \,,
\end{equation}
a barotropic equation of state. Remarkably, the contributions from the exchange pressure from Eq. (\ref{px}) and from the Dirac exchange potential are entirely equivalent, namely 
\begin{equation}
\nabla V_X = \frac{\nabla p_X}{n} \,,
\end{equation}
as shown from elementary algebra. This is the main result of this work. Finally, for ultra-cold ions obviously one has ${\bf P}_i = 0$.

\section{Conclusions}

A complete derivation of the Dirac exchange potential has been made, from newly introduced quantum plasma models taking into account the antisymmetry of the two-particle Wigner distribution function. The derivation from the exchange kinetic theory is quite different from the original derivation by Dirac, which is based on the Thomas-Fermi atomic model \cite{Dirac}. 

The simpler kinetic equation (\ref{e2}) is more amenable to nonlinear and numerical analysis, in comparison with Eq. (\ref{e1}), considered in the literature. Even more simplicity appears in the hydrodynamic modeling, which however needs an equation of state for closure. A local completely degenerate Fermi equilibrium shows a pressure term entirely equivalent to Dirac's exchange potential. Although evident from the start, it worth to mention that a classical, locally maxwellian equilibrium would gives something else \cite{h2021} than the Dirac expression. 

The approximations made, restrict the results to collisionless, semiclassical and completely degenerate plasmas. The domain of validity include real, dense fully degenerate and cold plasma systems, as identified in Fig. 1. Nevertheless, the procedure can be in principle extended to cover a finite-temperature electron gas, or a systematic derivation of correlation interaction potentials. The later would need a first principle kinetic equation including correlation besides exchange effects. However, at least in terms of the usual DFT parametrization of the exchange-correlation potential, the correlation contribution is shown to be a small correction in comparison with the exchange contribution, in the case of ideal plasmas. Finally, relativistic effects would be a further ingredient, in even more general theories. 





\appendix
\section{Derivation of the semiclassical exchange kinetic equation}

Denoting the right-hand side of Eq. (\ref{e1}) by $I$ 
%
%
and expanding in a formal power series of $\hbar$, it becomes
\begin{eqnarray}
I &=& - \,\frac{e^2 \hbar}{8\pi\varepsilon_0} \frac{\partial}{\partial p_i} \int d^{3}\rho \,d^{3} q \,\frac{\rho_i}{\rho^3} \,e^{-i\boldsymbol{\rho}\cdot{\bf q}} \, f(+) f(-) \nonumber \\ &+& \frac{e^2 \hbar^2}{16\pi\varepsilon_0} \frac{\partial}{\partial p_i} \frac{\partial}{\partial x_j}\int d^{3}\rho \,d^{3} q \,\frac{\rho_i\rho_j}{\rho^3} \,e^{-i\boldsymbol{\rho}\cdot{\bf q}} \, f(+) f(-) \nonumber \\
&+& \frac{e^2 \hbar^3}{64\pi\varepsilon_0} \frac{\partial}{\partial p_i}\frac{\partial}{\partial p_j}  \int d^{3}\rho \,d^{3} q \,\frac{\rho_i \rho_k}{\rho^3} \,e^{-i\boldsymbol{\rho}\cdot{\bf q}} \times \Bigl[f(+)\,\frac{\partial^2 f(-)}{\partial x_j \partial x_k} 
-
f(-)\,\frac{\partial^2 f(+)}{\partial x_j \partial x_k} \nonumber \\ 
&-& \frac{\partial f(+)}{\partial x_j} \frac{\partial f(-)}{\partial x_k}
+ \frac{\partial f(-)}{\partial x_j} \frac{\partial f(+)}{\partial x_k}\Bigr]
+  \, {\cal O}(\hbar^4) \,,
\label{a1}
\end{eqnarray}
summation convention implied, where for simplicity the time-dependence was omitted since it is not relevant for this discussion and where now the shorthand  
\begin{equation}
f(\pm) = f\left({\bf x}, {\bf p} \pm \frac{\bf q}{2}\right) \,.
\end{equation}
is used. 


It happens that the terms proportional to $\hbar$ and $\hbar^3$ vanishes in Eq. (\ref{a1}) due to parity properties, as can be verified by means of the simultaneous change of  variables ${\bf q} \rightarrow - {\bf q}, \boldsymbol{\rho} \rightarrow - \boldsymbol{\rho}$. For the surviving term proportional to $\hbar^2$ we use 
\begin{equation}
\int d^{3}\rho \,\frac{\rho_i\rho_j}{\rho^3}\,e^{-i\boldsymbol{\rho}\cdot{\bf q}} = i\,\frac{\partial}{\partial q_i}\,\int d^{3}\rho \,\frac{\rho_j}{\rho^3}\,e^{-i\boldsymbol{\rho}\cdot{\bf q}} = 
\frac{4\pi}{q^2}\left(\delta_{ij} - \frac{2 q_i q_j}{q^2}\right) \,,
\end{equation}
as can be verified using a partial integration and the Fourier transform of the Coulomb potential, 
\begin{equation}
\int \frac{d^{3} \rho}{\rho}\,e^{-i\boldsymbol{\rho}\cdot{\bf q}} = \frac{4\pi}{q^2} \,.
\end{equation}
Equation (\ref{a1}) becomes 
\begin{equation}
I = \frac{e^2 \hbar^2}{4\,\varepsilon_0} \frac{\partial}{\partial p_i} \frac{\partial}{\partial x_j}\int \frac{d^{3} q}{q^2}\,\left(\delta_{ij} - \frac{2 q_i q_j}{q^2}\right) f(+) f(-) + \, {\cal O}(\hbar^4) \,.
\end{equation}
A few more simple calculations finally give Eq. (\ref{e2}). It is worth mentioning that expanding to higher orders the terms proportional to $\hbar^\nu$ where $\nu$ is an odd integer vanishes due to parity properties. Hence, actually one has a power series on $\hbar^2$, as expected from quantum perturbation theory in general. 


\acknowledgments
The author acknowledges the support by Con\-se\-lho Na\-cio\-nal de De\-sen\-vol\-vi\-men\-to Cien\-t\'{\i}\-fi\-co e Tec\-no\-l\'o\-gi\-co
(CNPq) and the help of Jackson Galv\~ao for one figure.  Data Availability Statement: the data that support the findings of this study are available from the corresponding author upon reasonable request.

\end{document}